\documentclass[10pt,a4paper,english,prb,nobibnotes,twocolumn,aps,superscriptaddress]{revtex4-1}
\usepackage[T1]{fontenc}
\usepackage[latin9]{inputenc}
\usepackage{array}
\usepackage{amstext}
\usepackage{graphicx}
\usepackage{esint}

\makeatletter

\providecommand{\tabularnewline}{\\}
\newcommand{\lyxdot}{.}

 \@ifundefined{textcolor}{}
 {%
   \definecolor{BLACK}{gray}{0}
   \definecolor{WHITE}{gray}{1}
   \definecolor{RED}{rgb}{1,0,0}
   \definecolor{GREEN}{rgb}{0,1,0}
   \definecolor{BLUE}{rgb}{0,0,1}
   \definecolor{CYAN}{cmyk}{1,0,0,0}
   \definecolor{MAGENTA}{cmyk}{0,1,0,0}
   \definecolor{YELLOW}{cmyk}{0,0,1,0}
 }

\usepackage{dcolumn}

\def\d{\mathrm{d}}

\makeatother

\usepackage{babel}

\begin{document}
\bibliographystyle{apsrev4-1}

\title{Accurate ionic forces and geometry optimisation in linear scaling
density-functional theory with local orbitals}

\author{Nicholas D. M. Hine}

\thanks{Author to whom correspondence should be sent}

\email{nicholas.hine@imperial.ac.uk}

\affiliation{The Thomas Young Centre for Theory and Simulation of Materials, Imperial
College London, London SW7 2AZ, UK}

\author{Mark Robinson}

\affiliation{Theory of Condensed Matter, Cavendish Laboratory, J. J. Thomson Avenue,
Cambridge CB3 0HE, UK}

\author{Peter D. Haynes}

\affiliation{The Thomas Young Centre for Theory and Simulation of Materials, Imperial
College London, London SW7 2AZ, UK}

\author{Chris-Kriton Skylaris}

\affiliation{School of Chemistry, University of Southampton, Southampton SO17
1BJ, UK}

\author{Mike C. Payne}

\affiliation{Theory of Condensed Matter, Cavendish Laboratory, J. J. Thomson Avenue,
Cambridge CB3 0HE, UK}

\author{Arash A. Mostofi}

\affiliation{The Thomas Young Centre for Theory and Simulation of Materials, Imperial
College London, London SW7 2AZ, UK}

\date{\today}
\begin{abstract}
Linear scaling methods for density-functional theory (DFT) simulations
are formulated in terms of localised orbitals in real-space, rather
than the delocalised eigenstates of conventional approaches. In local-orbital
methods, relative to conventional DFT, desirable properties can be
lost to some extent, such as the translational invariance of the total
energy of a system with respect to small displacements and the smoothness
of the potential energy surface. This has repercussions for calculating
accurate ionic forces and geometries. In this work we present results
from \textsc{onetep}, our linear scaling method based on localised
orbitals in real-space. The use of psinc functions for the underlying
basis set and on-the-fly optimisation of the localised orbitals results
in smooth potential energy surfaces that are consistent with ionic
forces calculated using the Hellmann-Feynman theorem. This enables
accurate geometry optimisation to be performed. Results for surface
reconstructions in silicon are presented, along with three example
systems demonstrating the performance of a quasi-Newton geometry optimisation
algorithm: an organic zwitterion, a point defect in an ionic crystal,
and a semiconductor nanostructure. 
\end{abstract}
\maketitle

\section{Introduction\label{sec:Introduction}}

Conventional methods for atomistic simulations based on density-functional
theory~\cite{Hohenberg-Kohn,Kohn-Sham} (DFT), such as the plane-wave
pseudopotential approach~\cite{MCP_review}, have had an immense
impact on the way in which material properties are studied. Their
reach has extended beyond condensed matter physics into materials
science, chemistry, earth sciences, biochemistry and biophysics. In
spite of their success, the system-size accessible to such techniques
is limited because the algorithms scale with the cube of the number
of atoms. The quest to bring to bear the predictive power of DFT calculations
on ever larger systems has resulted in much interest in developing
linear scaling methods for DFT simulations~\cite{Goedecker_review,yang-prl91,galli-prl92,li-prb93,ordejon-prb95,hernandez-prb95,fattebert-prb00,skylaris-prb02,liu-prb03,fattebert-prb06,takayama-prb06},
and there are now a number of linear scaling DFT codes available,
including \textsc{onetep}~\cite{skylaris-jcp05,haynes-pssbssp06,hine-cpc-2009},
\textsc{conquest}~\cite{bowler-pssbssp06,gillan-cpc07}, \textsc{siesta}~\cite{soler-jpm02},
\textsc{openmx}~\cite{ozaki-prb05}, and other codes designed for
large-scale simulations, such as \textsc{bigdft}~\cite{bigdft_url}
and \textsc{fhi-aims}~\cite{blum_ab_2009}. The ability to perform
total energy calculations in $\mathcal{O}(N)$ operations, where $N$
is the number of atoms, is only the first step toward solving real
scientific problems, as most applications require structural relaxation.
This means computation of the ionic forces, and as such force calculations
are implemented in most of the codes listed above \cite{Miyazaki2004,torralba-jctc09,Ordejon_1996}
using a variety of choices of basis set.

One of the main advantages of using a plane-wave basis is that the
basis set is independent of ionic positions, hence there are no Pulay
corrections~\cite{Pulay1969} to the forces. As a result, the prefactor
associated with calculating ionic forces is small and constitutes
a negligible fraction of the total computational time. With the algorithms
used in plane-wave pseudopotential (PWP) simulations, forces cost
$\mathcal{O}(N)$ operations per ion, and hence $\mathcal{O}(N^{2})$
operations in total. However, it is not immediately clear that the
advantages of the PWP method for evaluation of forces can be carried
over to the context of real-space linear-scaling methods, for two
reasons. Firstly, these methods must be formulated in terms of objects
localised in real-space, and the delocalised nature of plane waves
would make them unsuitable as a basis set. Secondly, if one combines
a basis set that is fixed in space with localisation constraints on
the localised functions which depend on the ion coordinates, then
as the ions move, the basis functions will move relative to the localisation
regions and edge points may move in and out of the regions. This may
result in potential energy surfaces (PES), mapped out by displacement
of the ions, that are less smooth than those obtained when the extended
Kohn-Sham orbitals of conventional DFT calculations are used. This
phenomenon leads to ionic forces that are not exactly consistent with
the PES, thereby limiting the accuracy and convergence rate of structural
relaxations.

The linear scaling approach we address here, \textsc{onetep}~\cite{skylaris-jcp05},
uses a localised basis set of psinc functions which can be shown to
be equivalent to plane-waves~\cite{mostofi-jcp03} and has comparable
systematic convergence, overcoming the first of the difficulties listed
above. In this work we investigate the effect of the second problem,
namely the accuracy of ionic forces and the smoothness of the PES,
and compare our results for a number of challenging cases with those
obtained using conventional cubic scaling plane-wave calculations.
The ionic forces have been implemented in a quasi-Newton geometry
optimisation scheme ~\cite{clark-zfk05} and we show results of structural
relaxation on the Si(001) surface and three further examples: an organic
zwitterion, a point defect in alumina, and a GaAs nanocrystal. We
then demonstrate the efficient scaling of these methods to very large
system sizes by demonstrating the application of the method to extended
DNA strands containing up to 17000 atoms.

In Section~\ref{sec:theory} the features of our method which result
in its effectiveness will be discussed briefly. In Section~\ref{sec:forces}
we demonstrate that, as a result of the minimisation procedure used
and the properties of the orthogonal psinc basis set that is employed,
only the Hellmann-Feynman force on each ion is required. In Section~\ref{sec:tests}
we demonstrate the convergence and consistency of these calculated
forces, and in Section~\ref{sec:apps} results from the application
of this method in the \textsc{onetep} code to realistic systems will
be presented, and in Section~\ref{sec:conclusions} conclusions will
be drawn.

\section{Theoretical Background}

\label{sec:theory}

Linear scaling methods exploit the {}``nearsightedness''~\cite{Kohn_nearsightedness,prodan-kohn2005}
inherent in quantum many body systems by exploiting the localisation
of Wannier functions~\cite{Kohn_1959,Cloizeaux-2,Nencieu1983,He-Vanderbilt}
or the single-particle density matrix~\cite{Cloizeaux,Ismail-Beigi}.
In \textsc{onetep} the density matrix is expressed in a separable
form originally suggested by McWeeny~\cite{McWeeny} and subsequently
by Hern{á}ndez \textit{et al.}~\cite{hernandez-prb95} in the context
of linear scaling calculations: \begin{equation}
\rho(\mathbf{r,r'})=\sum_{\alpha\beta}\phi_{\alpha}(\mathbf{r})K^{\alpha\beta}\phi_{\beta}^{\ast}(\mathbf{r'}),\label{eqn:dm}\end{equation}
 where $\{K^{\alpha\beta}\}$ are the elements of the density kernel~\cite{McWeeny}
and $\{\phi_{\alpha}\}$ are a set of atom-centred non-orthogonal
generalised Wannier functions~\cite{skylaris-prb02} (NGWFs). Linear
scaling is achieved by imposing spatial cut-offs for the range of
the density kernel and localisation radii of the NGWFs. In our procedure
we mimimise the total energy with respect to both the density kernel
and the NGWFs. The Brillouin zone is sampled at the $\Gamma$-point
only.

In order to optimise the NGWFs they must be represented in some basis.
The plane-waves of conventional DFT calculations have many desirable
properties: the kinetic energy operator is diagonal in momentum space;
quantities are switched efficiently between real space and momentum
space using fast-Fourier transforms; the completeness of the basis,
and hence the accuracy of one's calculation, is controlled systematically
with a single parameter; and, particularly relevant to this work,
the ionic forces are calculated by straightforward application of
the Hellmann-Feynman theorem~\cite{hellmann_1937,feynman_1939}.

The extended nature of plane-waves, however, would appear to make
them unsuitable for describing the real-space localised orbitals used
in linear scaling methods. In spite of this, \textsc{onetep} is a
linear scaling method based on a plane-wave basis set which overcomes
the above difficulty and is able to achieve the same accuracy~\cite{skylaris-pssbssp06,haynes-cpl06}
and convergence rate~\cite{mostofi-jcp03} as the conventional plane-wave
approach.

\textsc{onetep} uses a localised yet orthogonal basis of periodic
cardinal sine (psinc) functions (defined in Ref.~\onlinecite{mostofi-jcp03})
which are formed from a discrete sum of plane-waves. As such it retains
many of the desirable properties inherent to the conventional plane-wave
approach. The localised NGWFs that span the occupied subspace are
represented in terms of these psinc functions and are optimised {\em
in situ} during the calculation.

\section{Ionic Forces and Geometry Optimisation}

\label{sec:forces}

In the context of Kohn-Sham DFT, the total energy $E$ is a functional
of the electronic density $n(\mathbf{r})$, which is given by the
diagonal part of the density matrix of Eq.~(\ref{eqn:dm}): \begin{equation}
n(\mathbf{r})=2\sum_{\alpha\beta}\phi_{\alpha}(\mathbf{r})K^{\alpha\beta}\phi_{\beta}^{\ast}(\mathbf{r}),\label{eqn:density}\end{equation}
 where the factor of two takes into account spin degeneracy.

In \textsc{onetep} the NGWFs are represented in terms of the underlying
orthogonal psinc basis $\{D_{i}(\mathbf{r})\}$: \begin{equation}
\phi_{\alpha}(\mathbf{r})=\sum_{i\in\mathrm{LR(\alpha)}}c_{i\alpha}D_{i}(\mathbf{r}),\label{eqn:psinc-exp}\end{equation}
 where LR($\alpha$) is the spherical, atom-centred localisation region
of NGWF $\phi_{\alpha}$ and $\{c_{i\alpha}\}$ are its expansion
coefficients in the psinc basis. Note that the localisation regions
move with the atoms but the locations of the points of the psinc grid
are fixed in space. Overall, the total energy is variationally dependent
on the coefficients $\{c_{i\alpha}\}$ and the elements $\{K^{\alpha\beta}\}$
of the density kernel. In our minimisation scheme we optimise all
of these degrees of freedom~\cite{skylaris-prb02}.

The force on an ion at $\mathbf{R}_{\gamma}$ is given by the derivative
of the total energy with respect to the ionic position, 
\begin{widetext}
\begin{equation}
\mathbf{F}_{\gamma}=-\frac{\mathrm{d}E}{\mathrm{d}\mathbf{R}_{\gamma}}=-\frac{\partial E}{\partial\mathbf{R}_{\gamma}}-\sum_{\alpha\beta}\frac{\partial E}{\partial K^{\alpha\beta}}\frac{\mathrm{d}K^{\alpha\beta}}{\mathrm{d}\mathbf{R}_{\gamma}}-\sum_{\alpha}\int\frac{\delta E}{\delta\phi_{\alpha}(\mathbf{r})}\frac{\mathrm{d}\phi_{\alpha}(\mathbf{r})}{\mathrm{d}\mathbf{R}_{\gamma}}d^{3}r.\label{eqn:force1}\end{equation}
 
\end{widetext}
Using Eq.~(\ref{eqn:psinc-exp}) and the fact that the psinc basis
is fixed, i.e., independent of ionic position such that $\frac{\mathrm{d}D_{i}(\mathbf{r})}{\mathrm{d}\mathbf{R}_{\gamma}}=0$,
the last term in Eq.~(\ref{eqn:force1}) may be expressed as \begin{equation}
\sum_{\alpha}\sum_{i\in\mathrm{LR}(\alpha)}\frac{\partial E}{\partial c_{i\alpha}}\frac{\mathrm{d}c_{i\alpha}}{\mathrm{d}\mathbf{R}_{\gamma}}.\end{equation}
Although the localisation regions move with the atoms, this does not
have any effect on the analytic derivative $\frac{\partial E}{\partial\mathbf{R}_{\gamma}}$,
because for an infinitesimal change $\delta\mathbf{R}_{\gamma}$ the
set of psinc points inside the localisation region, $i\in LR(\alpha)$,
does not change. 

At the end of the electronic minimisation, if we can assume that the
total energy is at a minimum with respect to the degrees of freedom
of the density, then we will satisfy the conditions \begin{equation}
\frac{\partial E}{\partial K^{\alpha\beta}}=0\ \ \text{and}\ \ \frac{\partial E}{\partial c_{i\alpha}}=0,\end{equation}
and we are on the Born-Oppenheimer surface for the given ionic configuration.
Under these conditions, the second and third terms in Eq.~(\ref{eqn:force1})
vanish, leaving only the Hellmann-Feynman force \begin{equation}
\mathbf{F}_{\gamma}=-\frac{\partial E}{\partial\mathbf{R}_{\gamma}},\label{eqn:force2}\end{equation}
which can be calculated in much the same spirit as the components
of the total energy itself, using our {}``fast Fourier transform
(FFT) box'' technique~\cite{skylaris-cpc01,mostofi-cpc02} to switch
quantities efficiently between real and reciprocal space.

The only components of the total energy with an explicit dependence
on $\mathbf{R}_{\gamma}$ are the ion-ion and electron-ion terms.
With nonlocal ionic pseudopotentials, there are both local and nonlocal
contributions to the latter. Written in Kleinman-Bylander form~\cite{Kleinman-Bylander},
the nonlocal pseudopotential energy is given by \begin{equation}
E_{\mathrm{nl}}=\sum_{\alpha\beta}\sum_{i}\langle\phi_{\alpha}|\chi_{i}\rangle D_{i}\langle\chi_{i}|\phi_{\beta}\rangle K^{\beta\alpha}\;,\end{equation}
where $\chi_{i}$ is the $i$-th projector, the sum over $i$ runs
over all the projectors on all atoms, and $D_{i}$ is its Kleinman-Bylander
energy. The nonlocal contribution to the force on atom $\gamma$ is
then \begin{eqnarray}
\mathbf{F}_{\gamma}^{\mathrm{nl}}=-\frac{\partial E_{\mathrm{nl}}}{\partial\mathbf{R}_{\gamma}} & = & -\sum_{\alpha\beta}\sum_{i}\Bigg[\Big<\phi_{\alpha}|\frac{\partial\chi_{i}}{\partial\mathbf{R}_{\gamma}}\Big>D_{i}\langle\chi_{i}|\phi_{\beta}\rangle\nonumber \\
 &  & \qquad-\langle\phi_{\alpha}|\chi_{i}\rangle D_{i}\Big<\frac{\partial\chi_{i}}{\partial\mathbf{R}_{\gamma}}|\phi_{\beta}\Big>\Bigg]K^{\beta\alpha}\;,\end{eqnarray}
where the sum over projectors runs only over those projectors on atom
$\gamma$. Since the projectors are only nonzero within the core region
of each ion and the NGWFs are strictly localised, projector-NGWF overlap
matrices $\langle\chi_{i}|\phi_{\beta}\rangle$ are highly sparse,
and evaluation of the overlaps is performed within the {}``FFT box''
approximation~\cite{skylaris-cpc01,mostofi-cpc02}. The nonlocal
contribution to the energy and all ionic forces can therefore be calculated
in $O(N)$ computational effort.

For the long-ranged Coulombic ion-ion and electron-ion terms, there
are ways to reformulate the Ewald method so that they scales as $O(N\ln N)$
with suitable approximations involving transferring the point charges
to a grid \cite{essmann_smooth_1995}. These methods are routinely
employed in classical MD codes, but are not easily amenable to high-accuracy
$O(N)$ methods in the size range considered here and come at a cost
in accuracy. Fortunately, the evaluation of these terms is nonetheless
computationally straightforward: within the Ewald approach, the ion-ion
term is \begin{equation}
\mathbf{F}_{\gamma}^{\mathrm{ew}}=-\frac{\partial E_{\mathrm{ew}}}{\partial\mathbf{R}_{\gamma}}\;,\end{equation}
which can be evaluated easily by standard techniques and can be made
to scale as $O(N^{3/2})$ or better\cite{perram_algorithm_1988},
with suitably chosen parameters. The local ionic pseudopotential contribution
is most easily evaluated in reciprocal space, as \begin{equation}
\mathbf{F}_{\gamma}^{\mathrm{loc}}=-\frac{\partial E_{\mathrm{loc}}}{\partial\mathbf{R}_{\gamma}}=\sum_{\mathbf{G}}i\mathbf{G}e^{-i\mathbf{G}.\mathbf{R}_{\gamma}}V_{\gamma}^{\mathrm{\mathrm{loc}}}(\mathbf{G})n^{*}(\mathbf{G})\;,\end{equation}
where $V_{\gamma}^{\mathrm{loc}}$ is the local pseudopotential of
atom $\gamma$. This is also relatively straightforward to compute
but clearly asymptotically involves $O(N^{2})$ computational effort
in this formulation, since the number of $\mathbf{G}$-vectors in
the simulation cell scales as $O(N)$ and the summation must be performed
for all $N$ atoms. As for the Ewald approach, it is possible to reformulate
this as an $O(N\ln N)$ algorithm \cite{choly_fast_2003,hung_accurate_2009},
but as will be seen in Section~\ref{sec:tests}, the $O(N^{2})$
contribution to the forces calculation has a small prefactor compared
to the $O(N)$ evaluation of the total energy, and does not become
problematic until the very largest system sizes currently encountered
in linear-scaling DFT calculations. To go further, one could alternatively
use fast multipole methods to reduce the scaling\cite{greengard_fast_1987}.

Finally, in systems with nonlinear core corrections to the exchange-correlation
energy \cite{louie_nonlinear_1982,dal_corso_ab_1993}, there is an
additional contribution to the force due to the fact that the core
density moves with the ion. This is also most easily evaluated in
reciprocal space, with a similar prefactor to the local potential
term: \begin{equation}
\mathbf{F}_{\gamma}^{\mathrm{nlcc}}=-\frac{\partial E_{\mathrm{nlcc}}}{\partial\mathbf{R}_{\gamma}}=\sum_{\mathbf{G}}i\mathbf{G}e^{-i\mathbf{G}.\mathbf{R}_{\gamma}}n_{\gamma}^{\mathrm{\mathrm{c}}}(\mathbf{G})V_{\mathrm{xc}}^{*}(\mathbf{G})\;.\end{equation}

Eq.~(\ref{eqn:force2}) is correct in the limit in which no localisation
constraints are imposed on the NGWFs. With localisation constraints,
the translational invariance of the system with respect to the grid
of psinc basis functions is broken~\cite{bernholc-ijqc97}, coined
the {}``egg-box'' effect~\cite{artacho-jpm08}, which introduces
an error in the force that is, in general, difficult to calculate
explicitly but which may be controlled by decreasing the grid spacing
or increasing the radius of the localisation regions~\cite{fattebert-cpc04}.
This phenomenon is related to the fact that the underlying basis of
psinc functions is fixed with respect to the ions while the LRs are
atom-centred and therefore move with the atoms, resulting in each
NGWF having a non-equivalent representation in the psinc basis depending
on its exact position with respect to the grid of psinc functions.
As will be demonstrated in Section~\ref{sec:tests}, forces calculated
in \textsc{onetep} according to Eq.~(\ref{eqn:force2}) are already
very accurate, even for weakly bonded systems, and have been implemented
in a quasi-Newton geometry optimisation scheme~\cite{Pfrommer1997}
based on the Broyden-Fletcher-Goldfarb-Shanno (BFGS) algorithm (see,
e.g., Ref.~\onlinecite{nocedal-optimization}).

\section{Preliminary Tests\label{sec:tests}}

\subsection{Convergence in molecular systems\label{sub:convergence_molecules}}

We start with preliminary calculations to examine issues of calculating
individual forces. First we present two very simple, small-scale test
cases: (i) a symmetric stretch of a carbon dioxide molecule, and (ii)
a hydrogen bond in a water dimer. In order to demonstrate the accuracy
of potential energy surfaces obtained from \textsc{onetep}, we compare
with equivalent calculations with the \textsc{castep}~\cite{clark-zfk05}
plane-wave pseudopotential code. In all comparisons we use identical
norm-conserving pseudopotentials~\cite{Troullier_Martins} in Kleinman-Bylander
separable form~\cite{Kleinman-Bylander}, the same local-density
approximation~\cite{Perdew-Zunger} for the exchange-correlation
functional, and $\Gamma$-point sampling of the Brillouin zone.

\begin{figure}[ht]
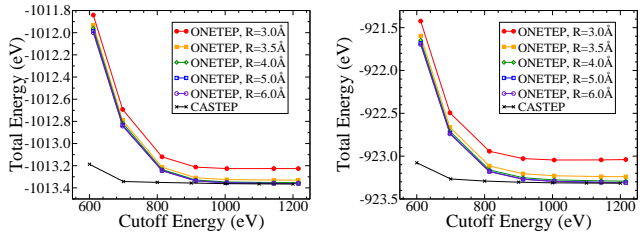

 \includegraphics[clip,width=0.46\columnwidth]{co2\lyxdot te_conv}$\quad$\includegraphics[clip,width=0.46\columnwidth]{h2o\lyxdot dim\lyxdot te_conv}
\caption{\label{fig:te_convergence}Total energy as calculated by \textsc{onetep}
and \textsc{castep} at a fixed geometry as a function of cutoff energy
$E_{c}$ and NGWF truncation radius $R_{\phi}$, for the CO$_{2}$
molecule (left) and an H$_{2}$O dimer (right). Convergence properties
are domininated by the oxygen pseudopotential and so are broadly the
same for the two systems. The results for 4.0\textbf{~}\AA{} spheres
to 6.0\textbf{~}\AA{} spheres agree to high accuracy, particularly
at a cutoff energy of 900~eV or above, indicating convergence at
around 4.0\textbf{~}\AA{}.}

\end{figure}

\begin{figure}[ht]
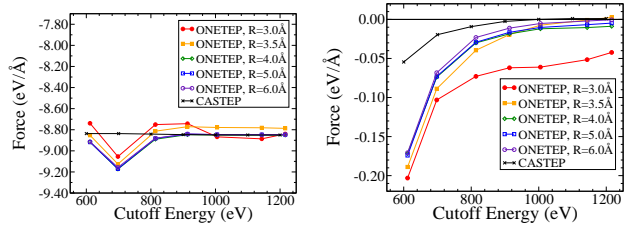

\includegraphics[width=0.44\columnwidth]{co2\lyxdot f_conv}$\quad$\includegraphics[width=0.46\columnwidth]{h2o\lyxdot dim\lyxdot f_conv}
\caption{\label{fig:f_convergence}Convergence properties of a single force
component calculated using \textsc{onetep} and \textsc{castep} at
a fixed, off-equilibrium geometry as a function of cutoff energy $E_{c}$
and NGWF truncation radius $R_{\phi}$, for the CO$_{2}$ molecule
(left) and an H$_{2}$O dimer (right). Force shown is the force acting
on an O atom along the bond axis for CO$_{2}$, (hence a very strong
force), and the force on the hydrogen atom not involved in the dimer-dimer
h-bond (hence a very weak force, near-equilibrium). Forces converge
with both $R_{\phi}$ and $E_{c}$, but care must be taken to accurately
converge with respect to both quantities simultaneously.}

\end{figure}

We must first investigate the convergence of calculated quantities
with respect to the basis size in the two methods. In \textsc{onetep}
we can systematically control the convergence by decreasing the grid
spacing (which corresponds to a plane-wave cutoff $E_{c}$) and increasing
the localisation radii $R_{\phi}$ of the NGWFs. In \textsc{castep}
we can vary the plane wave cutoff $E_{c}$. In Fig.~\ref{fig:te_convergence}
we examine the convergence of the total energy of the two molecular
systems with respect to these quantities, while in Fig.~\ref{fig:f_convergence}
we examine the convergence of force components. It is observed, as
previously noted \cite{skylaris-jpm05}, that total energy convergence
with grid spacing is slower in \textsc{onetep} than in plane-wave
methods such \textsc{castep}. This is due to the greater influence
of the so-called `egg-box' effect in the former (the variation in
energy with uniform translation of the atoms with respect to the grid),
which results from NGWF truncation to points within a sphere on a
regular underlying grid. However, both methods converge asymptotically
to the same value to a high precision. The \textsc{onetep} forces
converge non-monotonically with both $E_{c}$ and $R_{\phi}$ to eventual
good agreement with the \textsc{castep} equivalents, at a radius not
significantly greater than would be required for tolerable convergence
of the total energy (around 4.0~\AA{} in this case). It is to be
noted that at a fixed, under-converged value of $E_{c}$, convergence
with $R_{\phi}$ is to an incorrect value, while at fixed, under-converged
$R_{\phi}$, convergence with $E_{c}$ may be erratic or tend to an
incorrect value. In particular, very small localisation regions result
in highly inaccurate forces. This emphasises the importance of converging
with respect to both parameters simultaneously for accurate results.
Finally, as with for plane-wave codes, it should be noted that accurate
forces will often require higher convergence parameters than accurate
energy differences.

\begin{figure}[ht]
\includegraphics[width=0.64\columnwidth]{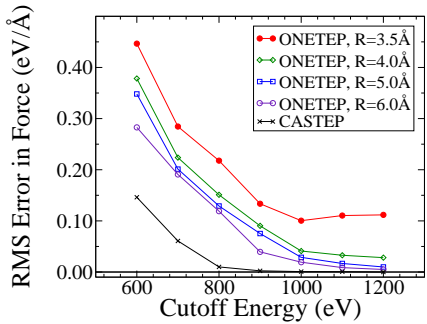} \caption{\label{fig:f_convergence-CHAPS}Convergence properties of the RMS
error in the \textsc{onetep} and \textsc{castep} forces at a fixed,
off-equilibrium geometry as a function of cutoff energy $E_{c}$ and
NGWF truncation radius $R_{\phi}$, for the CHAPS molecule (see text).
The quantity plotted is the RMS difference between the calculated
result for a given $E_{c}$ and $R_{\phi}$ and the calculated result
for $E_{c}=1200$eV in \textsc{castep}, which is taken to be the converged
result. We see that with respect to both $R_{\phi}$ and $E_{c}$
it is possible to obtain systematic convergence to the plane-wave
result.}

\end{figure}

To demonstrate that the convergence properties demonstrated here are
applicable to larger systems, we examine the difference between the
calculated forces in \textsc{onetep} and \textsc{castep} for a larger
molecule. We employ the organic zwitterionic detergent molecule, 3-{[}(3-Cholamidopropyl)dimethylammonio{]}-1-propanesulfonate,
or `CHAPS' as a particularly challenging, worst-case test system,
due to the considerable charge separation and long-ranged forces encountered.
In Figure \ref{fig:f_convergence-CHAPS} we plot the RMS error of
all the calculated forces, and observe that they can be systematically
converged with respect to local orbital radius and cutoff energy.

\subsection{Consistency in molecular systems\label{sub:CO2-and-H2O-1}}

We now examine the consistency of the forces and the energy by calculating
the full binding curve for the CO$_{2}$ molecule and the H$_{2}$O
molecular dimer. For the \textsc{castep} calculation a plane-wave
energy cut-off energy of 1100~eV was used for the wavefunctions,
while for \textsc{onetep} a grid-spacing of 0.227~\AA{}\ (equivalent
to a plane-wave energy cut-off of 1121~eV) and four NGWFs on each
atom, each with a localisation radius of 4.0~\AA{}, were used. Examination
of Figs.~\ref{fig:te_convergence} and \ref{fig:f_convergence} suggest
these cutoffs will produce results accurate to within around 10~meV
for the total energy and 0.01~eV/\AA{}\ for the forces in these
two systems.

\begin{figure}[ht]
 \includegraphics[clip,width=0.9\columnwidth]{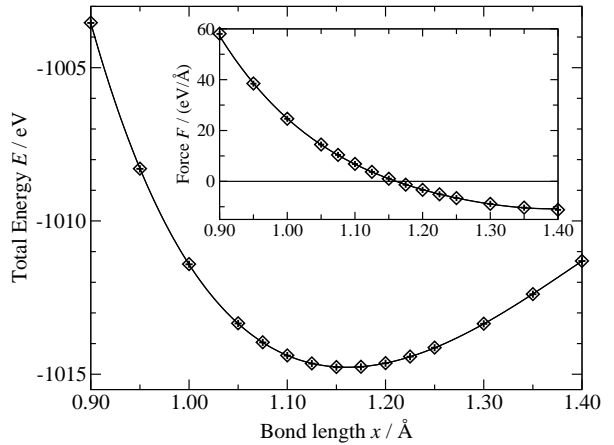}
\caption{\label{fig:co2} Total energy as a function of C--O bond-length for
a carbon dioxide molecule subjected to a symmetric stretch. \textsc{castep}
(plus symbols); \textsc{onetep} (open diamonds). The curve shows a
polynomial fit $E(x)$ to the data points. On the scale of this plot
the data are indistinguishable. Inset: the Hellmann-Feynman force
on each oxygen atom. The curve shows the numerical force $F(x)\equiv-\frac{1}{2}\frac{\d E}{\d x}$.}

\end{figure}

In Fig.~\ref{fig:co2} the variation of the total energy of a carbon
dioxide molecule with respect to the C--O bond length is shown, as
calculated with \textsc{castep} (plus symbols) and \textsc{onetep}
(open diamonds). The curve drawn is a polynomial fit $E(x)$ to the
data points. As can be seen, the results are indistinguishable. The
inset to Fig.~\ref{fig:co2} shows the Hellmann-Feynman force, calculated
according to Eq.~(\ref{eqn:force2}), in \textsc{castep} (plus symbols)
and \textsc{onetep} (open diamonds). Again the agreement is excellent.
The curve shown in the inset is the analytic derivative $F(x)\equiv-\frac{1}{2}\frac{\d E}{\d x}$
of the polynomial fit to the energy data points. A discrepancy between
this curve and the calculated forces would indicate an inconsistency
between the PES and the Hellmann-Feynman forces. We see that there
is no inconsistency and that any errors introduced by imposing localisation
constraints on the NGWFs are negligible. Quantitative comparison of
the two approaches shows that the fractional differences in the equilibrium
bond length and the curvature of the PES at the minimum, respectively,
are less than 0.1\% and 0.2\%. The discrepancy between the equilibrium
bond length as predicted by the numerical force (given by the $x$-intercept
of $F(x)$) and by the Hellmann-Feynman force (given by a polynomial
fit to the force data points) is less than 0.1\% for both approaches.

\begin{figure}[ht]
 \includegraphics[width=0.9\columnwidth]{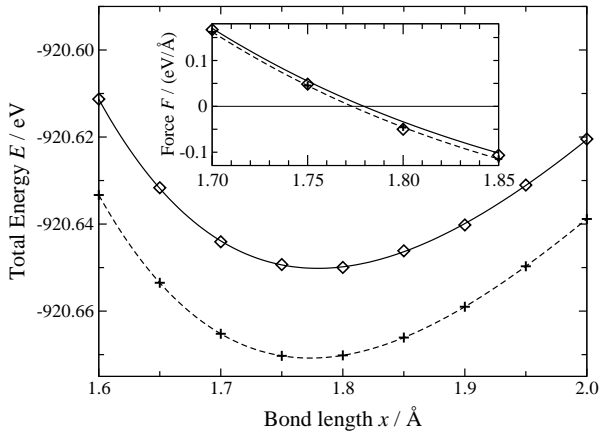}
\caption{\label{fig:water-dimer} Interaction potential of a water dimer. \textsc{castep}
(plus symbols); \textsc{onetep} (open diamonds). The curves show polynomial
fits $E(x)$ to the data points. Inset: the Hellmann-Feynman force
on each water molecule. The curves show the numerical force $F(x)\equiv-\frac{\d E}{\d x}$
for \textsc{castep} (dashed line) and \textsc{onetep} (solid line).}

\end{figure}

We turn now to the more challenging case of the water dimer. In Fig.~\ref{fig:water-dimer}
the total energy as a function of the length of the hydrogen bond
is shown. This is a much more sensitive test, as the potential well
associated with the hydrogen bond is shallower and the forces weaker
by two orders of magnitude than in the case of the strong covalent
bonds in carbon dioxide.

For the \textsc{castep} calculation a plane-wave cut-off energy of
1200~eV was used. For \textsc{onetep}, a grid-spacing of 0.214~\AA{}
(equivalent to a plane-wave energy cut-off of 1261~eV) was used;
each hydrogen and oxygen atom had one and four NGWFs, respectively,
all of radius 4.0~\AA{}.

From Fig.~\ref{fig:water-dimer} it may be seen that the total energies
of the two approaches are within 20~meV of each other. The predicted
equilibrium bond length and the curvature of the PES at the minimum
agree to within 0.3\% and 4.6\%, respectively, well within the variations
associated with using different exchange and correlation functionals.
The effect of localisation constraints, however, is now apparent.
The inset shows that the Hellmann-Feynman forces in \textsc{onetep}
do not coincide perfectly with the derivative of the fit to the total
energy. Nevertheless, the discrepancy is small: at the equilibrium
bond length the error, defined as the fractional difference between
the $x$-intercept of the numerical force (solid line, inset of Fig.~\ref{fig:water-dimer})
and that of a fit (not shown) to the Hellmann-Feynman forces (open
diamonds, inset of Fig.~\ref{fig:water-dimer}) is only 0.3\%, or
0.005~\AA{}. Note that the forces in \textsc{onetep} agree with the
forces in \textsc{castep} even more accurately than the derivative
of the fit to the \textsc{onetep} total energy agrees with the corresponding
forces. This correlates well with the observed insensitivity of the
calculated forces to the effect of truncation of the local orbitals.

\subsection{Convergence and Consistency in Bulk Systems}

The above tests for molecules demonstrate the basic applicability
of the methods for evaluation of forces within the context of this
particular local-orbital method, but do not test their accuracy in
the more challenging conditions of a solid, with the constraints necessary
for linear-scaling. In particular, in a solid, convergence with local
orbital radius can present greater difficulties. To demonstrate the
convergence behaviour in solids, we simulate a block of bulk silicon
(diamond structure) subject to random distortions of the atomic coordinates
about their equilibrium positions and compare the calculated \textsc{onetep}
forces on these displaced atoms with those calculated in \textsc{castep}. 

We begin with a supercell consisting of $5\times5\times5$ times the
8-atom cubic unit cell of the bulk silicon with lattice parameter
$a=5.4$\textbf{~}\AA{}, giving 1000 atoms. We then generate a set
of realisations of random disorder by displacing each atom according
to a uniform random distribution with a given amplitude. These systems
are not intended to be physically meaningful apart from as a test
of the accuracy of the calculated forces, but are approximately representative
of a snapshot of the system at elevated temperature. Calculations
were performed with\textbf{ }\textsc{onetep}\textbf{ }for\textbf{
$R_{\phi}=4.23$}~\AA{}, \textbf{$R_{\phi}=4.76$}~\AA{} and \textbf{$R_{\phi}=5.29$}~\AA{}.
A fixed psinc spacing of $0.256$\textbf{~}\AA{} corresponding to
a plane-wave cutoff of $E_{c}=883$eV, which was verified to give
good convergence of both total energies and forces in \textsc{castep}
was used in both codes. In the\textsc{ onetep}\textbf{ }calculations,
a kernel cutoff greater than the supercell size was employed, meaning
all elements of the density kernel were nonzero, and optimised using
the LNV energy minimisation scheme \cite{li-prb93}. The \textsc{castep}
calculations were performed on identical 1000-atom cells with the
same plane-wave cutoff. We employed the Local Density Approximation
for exchange and correlation. For many systems it is possible to obtain
plane-wave accuracy using only as many NGWFs as valence orbitals~\cite{skylaris-jpm05}.
However previous \textsc{onetep} calculations on bulk silicon \cite{skylaris-jpm07}
have reported that nine NGWFs per silicon atom are required to achieve
plane-wave accuracy, and this prescription was followed here.

Table \ref{tab:Si_bulk_F_conv} shows the convergence of the forces
with respect to the local orbital radius. We see that by \textbf{$R_{\phi}=4.23$}~\AA{}
results are already in reasonable agreement with equivalent plane-wave
results, with an RMS deviation increasing from 0.001~eV/\AA{} to
0.004~eV/\AA{} as the disorder magnitude $\Delta$ increases. For
larger radii the results improve, though the extent to which they
can agree with the \textsc{castep} results is limited by the `egg-box'
effect resulting from the underlying grid spacing. By $\Delta=0.5a_{0}$,
the system is some way off equilibrium, with an RMS force of around
2.3~eV/\AA{}, but the precision of the agreement with plane-wave
results is maintained.

\begin{table*}[ht]
\begin{ruledtabular}
\begin{tabular}{rr>{\raggedleft}p{0.16\columnwidth}rr>{\raggedleft}p{0.16\columnwidth}}
$\Delta$ ($a_{0}$) & $R_{\phi}$ ($a_{0}$) & $\phantom{-\mathbf{F}_{\text{CAS}}}|\mathbf{F}|_{\text{max}}$ & $|\mathbf{F}-\mathbf{F}_{\text{CAS}}|_{\text{max}}$ & $\phantom{-\mathbf{F}_{\text{CAS}}}|\mathbf{F}|_{\text{rms}}$ & $|\mathbf{F}-\mathbf{F}_{\text{CAS}}|_{\text{rms}}$\tabularnewline
\hline
0.00 & 8.0 & 0.00135 & 0.00135 & 0.00096 & 0.00096\tabularnewline
0.00 & 9.0 & 0.00020 & 0.00020 & 0.00014 & 0.00014\tabularnewline
0.00 & 10.0 & 0.00011 & 0.00011 & 0.00008 & 0.00008\tabularnewline
\hline
0.05 & 8.0 & 0.40748 & 0.00437 & 0.22083 & 0.00193\tabularnewline
0.05 & 9.0 & 0.40772 & 0.00357 & 0.22095 & 0.00175\tabularnewline
0.05 & 10.0 & 0.40779 & 0.00354 & 0.22091 & 0.00169\tabularnewline
\hline
0.50 & 8.0 & 6.1615 & 0.00870 & 2.22984 & 0.00414\tabularnewline
0.50 & 9.0 & 6.1648 & 0.00450 & 2.30108 & 0.00208\tabularnewline
0.50 & 10.0 & 6.1620 & 0.00739 & 2.29999 & 0.00270\tabularnewline
\end{tabular}
\end{ruledtabular}
\caption{\label{tab:Si_bulk_F_conv}Convergence with NGWF radius of atomic
forces (eV/\AA{}) for a 1000-atom system of bulk Si subject to random
displacements of $\Delta=\{0,0.05,0.5\}a_{0}$, where $a_{0}=0.529$\textbf{~}\AA{}.
$\mathbf{F}$ is the force calculated in \textsc{onetep} and $\mathbf{F}_{\text{CAS}}$
its equivalent calculated in \textsc{castep.} Maximum and root mean
square forces, plus corresponding values for the maximum and RMS deviation
of forces from the \textsc{castep} result are shown for each of $R_{\phi}=\{8,9,10\}a_{0}$.
Note that for $\Delta=0$, the \textsc{castep }forces are all zero
by symmetry, and the deviation of the\textsc{ onetep} results is solely
due to the `egg-box' effect and the influence of NGWF truncation.}

\end{table*}

\section{Applications\label{sec:apps}}

\subsection{Si Surface Reconstructions}

We now report results of a realistic application of geometry optimisation,
using our quasi-Newton scheme, on a Si(001) surface, comparing again
with \textsc{castep}. Calculations were performed within the local-density
approximation in a 8$\times$8 supercell consisting of nine atomic
layers of silicon atoms in which the bottom layer of atoms was hydrogen
passivated (a total of 640 atoms). A fixed lattice constant of 5.43~\AA{}\ was
used, resulting in a supercell of dimensions 30.713~\AA{}\ $\times$
30.713~\AA{}\ in the plane of the surface. The size of the supercell
in the perpendicular direction was 25.595~\AA{}, providing a vacuum
gap of 12.9~\AA{}\ between adjacent periodic replicas, though this
varies slightly during the calculation. For these surface calculations,
the same grid spacing of 0.256~\AA{}\ was used, and localisation
radii of 4.0~\AA{} were chosen.

The passivating hydrogen atoms were constrained to lie vertically
below the bottom layer of silicon atoms which were fixed to their
bulk positions. Surface atoms were given small initial random displacements
to break symmetry so that they could dimerise. Symmetry was imposed
so that the surface could only form a p(2$\times$1) reconstruction.
This allows a direct comparison with a \textsc{castep} calculation
on a 2$\times$1 surface supercell comprised of 18 silicon atoms and
two passivating hydrogen atoms. A plane-wave cut-off of 883~eV was
used for the wavefunctions and the Brillouin zone was sampled using
an evenly spaced grid consisting of 4$\times$8 k-points in \textsc{castep}.
The same pseudopotentials were used for both \textsc{onetep} and \textsc{castep}
calculations.

\begin{figure}[ht]
 \includegraphics[width=0.9\columnwidth]{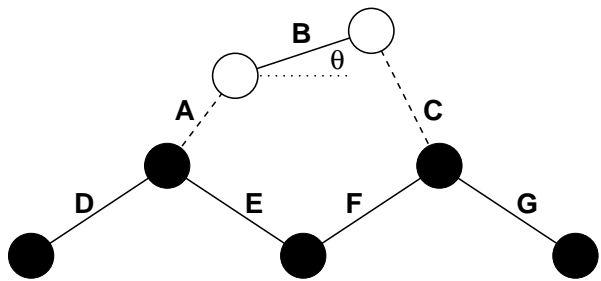} \caption{\label{fig:si_dimer} Schematic of the reconstructed Si(001) surface.
Bonds indicated by dashed lines do not lie in the plane of the diagram.
The surface atoms (indicated by white circles) pair up to form dimers,
that then buckle out of the plane of the surface. For the purposes
of this work the bonds have been labelled alphabetically and the buckling
angle denoted by $\theta$.}

\end{figure}

\begin{table*}[ht]
\begin{ruledtabular}
\begin{tabular}{lllllllll}
 & \multicolumn{1}{c}{A} & \multicolumn{1}{c}{B} & \multicolumn{1}{c}{C} & \multicolumn{1}{c}{D} & \multicolumn{1}{c}{E} & \multicolumn{1}{c}{F} & \multicolumn{1}{c}{G} & \multicolumn{1}{c}{$\theta$}\tabularnewline
\hline 
\textsc{onetep}  & 2.306  & 2.272  & 2.364  & 2.362  & 2.398  & 2.333  & 2.364  & 17.9$^{\circ}$\tabularnewline
\textsc{castep}  & 2.313  & 2.274  & 2.371  & 2.378  & 2.403  & 2.338  & 2.380  & 17.0$^{\circ}$ \tabularnewline
Ref \cite{ramstad-prb95}. & 2.29  & 2.26  & 2.34  & 2.35  & 2.38  & 2.33  & 2.35  & 18.3$^{\circ}$\tabularnewline
\end{tabular}
\end{ruledtabular}
\caption{\label{tab:si_results} Bond lengths (in Angstrom) and the buckling
angle $\theta$ as calculated by \textsc{onetep}, \textsc{castep}
and Ramstad {\em et al.}~\cite{ramstad-prb95}.}

\end{table*}

As expected surface atoms were observed to pair up to form dimers,
which then buckled out of the plane of the surface. The resulting
geometry is shown in Fig.~\ref{fig:si_dimer}. Bond lengths and buckling
angles are compared in Table~\ref{tab:si_results}. They compare
very well with all bond lengths lying within 0.02~\AA{}. The differences
in bond lengths lead to a slightly smaller buckling angle in \textsc{castep}
compared to \textsc{onetep}. The bond lengths also compare well with
those found in previous work by Ramstad \emph{et. al.}~\cite{ramstad-prb95}.
The shorter bond lengths found in that work can be attributed to the
use of different pseudopotentials.

Regarding the number of NGWFs/atom, we observed here that using only
four NGWFs per silicon, surface atoms did dimerise but the resulting
dimers failed to buckle. The flexibility afforded by nine NGWFs appears
to be required for the dimers to relax into the buckled geometry.

\subsection{Charge Redistribution}

\begin{figure*}[ht!]
 \includegraphics[width=0.5\columnwidth]{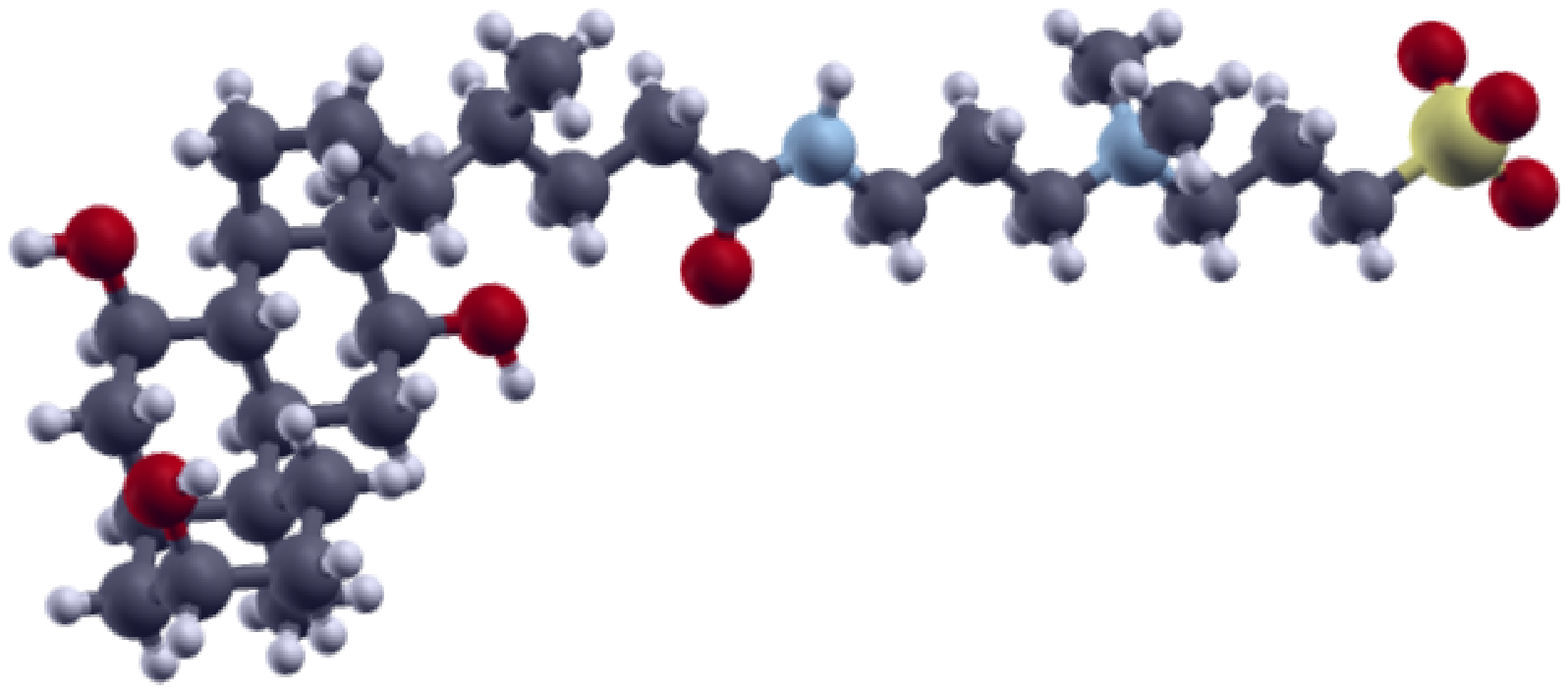} \includegraphics[width=0.5\columnwidth]{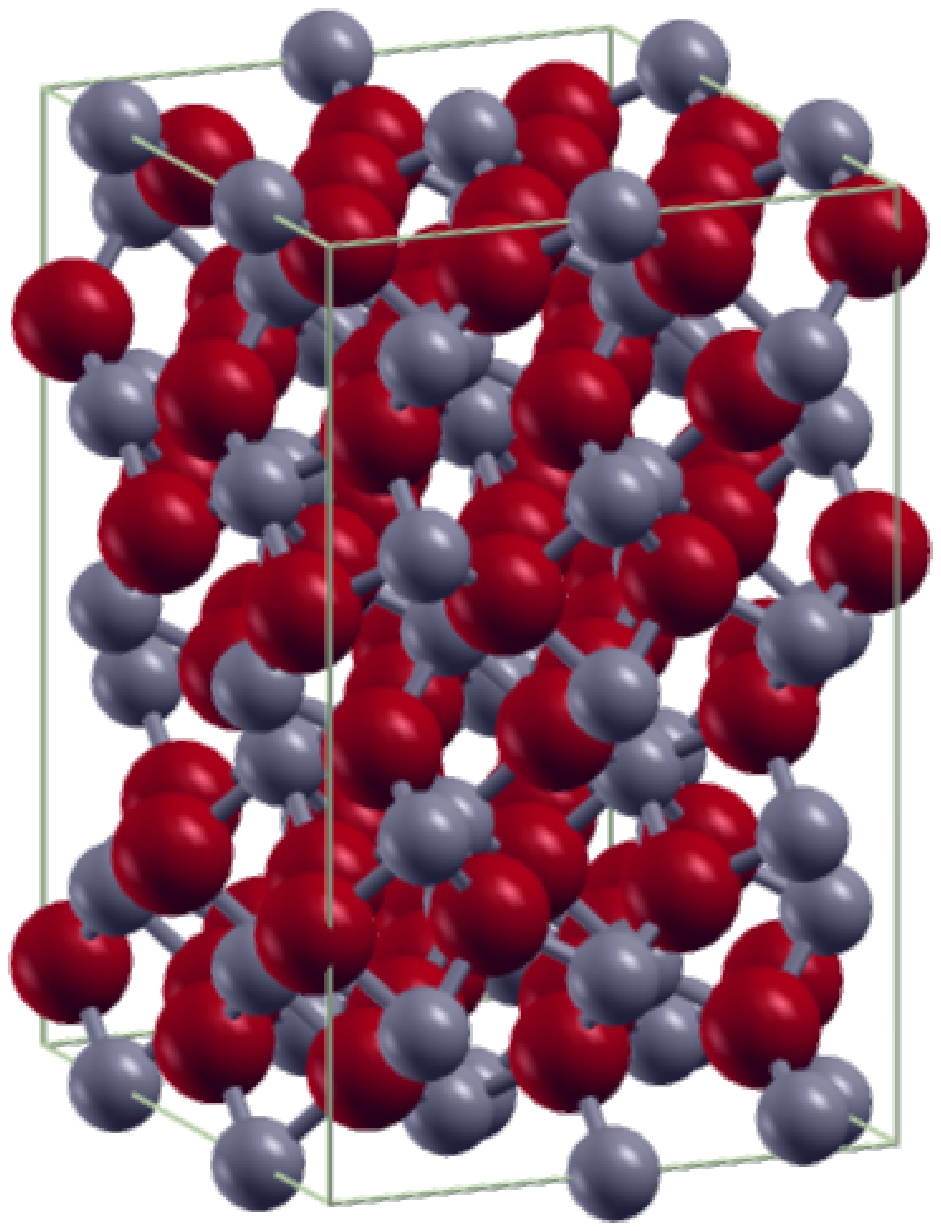}
\includegraphics[width=0.5\columnwidth]{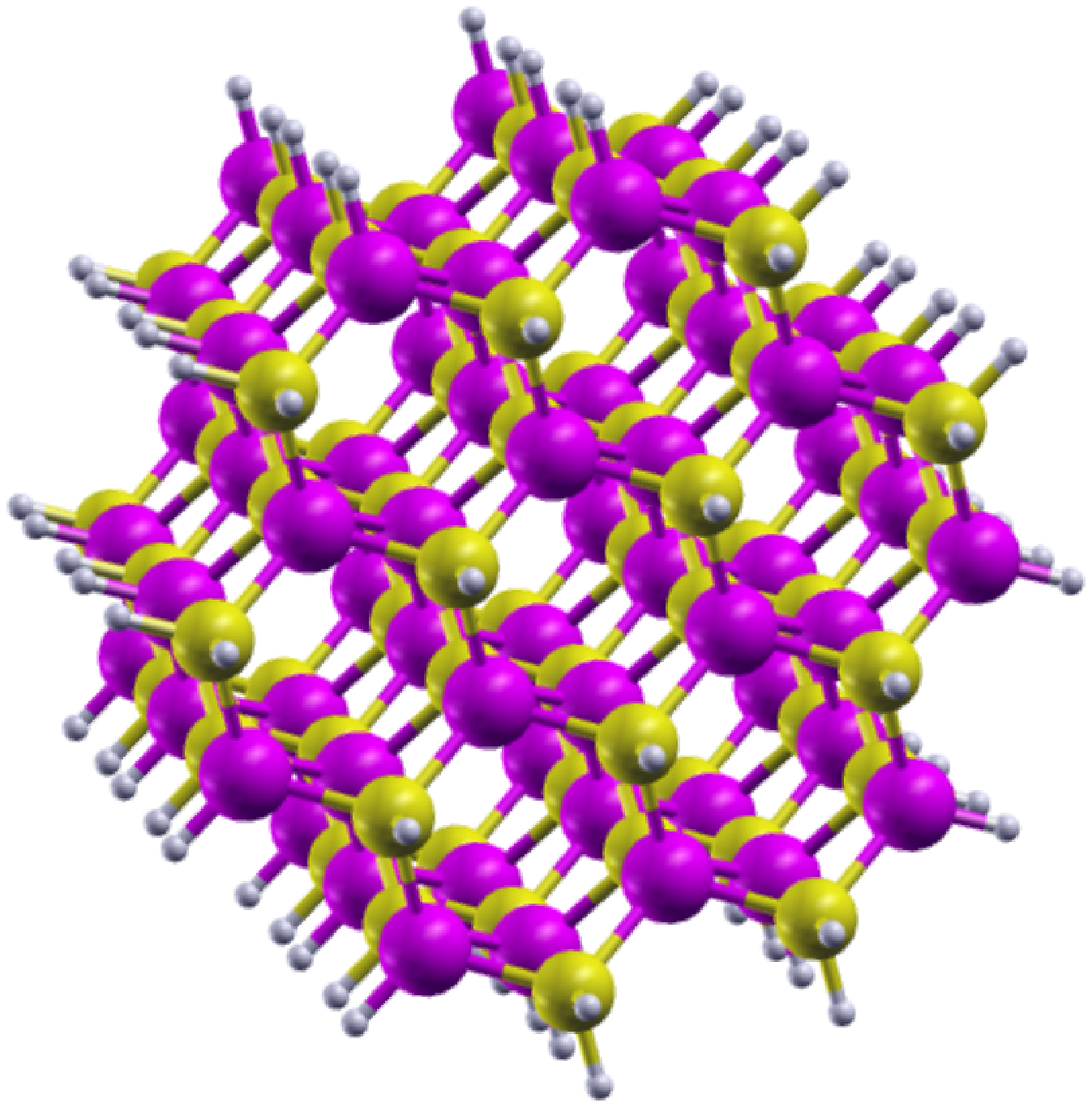} \caption{\label{fig:bfgs-conv-pics} Systems for which geometry optimisation
was performed with the BFGS algorithm in \textsc{onetep} for illustration
of convergence behaviour. Left: CHAPS molecule (100 atoms) Centre:
Al vacancy in $2\times2\times1$ supercell of $\alpha-$alumina (119
atoms) Right: H-terminated GaAs nanocrystal (204 atoms)}

\end{figure*}

Finally we report on geometry convergence tests for structural optimisation
in more complex systems exhibiting charge redistribution, to test
the performance of the algorithms. System (a) is the CHAPS molecule
referred to in Sec.~\ref{sub:convergence_molecules}. Its zwitterionic
nature leads to considerable charge separation and some relatively
long-ranged contributions to the forces, hence a challenging case
for geometry relaxation. We started from standard crystallographic
data~\cite{KleywegtJones}, with hydrogen atoms added to saturate
dangling bonds. The resulting molecule contains 100 atoms. System
(b) is a crystalline ceramic of 119 atoms, comprising a $2\times2\times1$
supercell of $\alpha-$alumina in the corundum structure, containing
one aluminium vacancy $\mathrm{V}_{\mathrm{Al}}^{-3}$ in charge state
$-3$ (such that neighbouring oxygens retain filled $p$-shells).
The starting configuration used for the relaxation was the optimised
bulk geometry before removal of the aluminium atom. In this ionic
system, containing a vacancy with a large net charge, there are again
considerable long-ranged relaxations. Finally, system (c) is a small
nanocrystal of wurtize-structure GaAs, a polar semiconductor. Starting
from the bulk wurtzite crystal structure optimised within DFT, the
nanocrystal is imagined to have been formed by cleaving to expose
{[}0001{]} faces on the two ends, corresponding to Ga and As layers,
respectively. There remains a net dipole moment parallel to the $c$-axis,
whose value depends on the geometry of the surfaces. The rod, comprising
204 atoms once dangling bonds are terminated with hydrogen, was simulated
inside a cubic simulation cell of side-length 45~\AA{}. These systems
are illustrated in Fig.~\ref{fig:bfgs-conv-pics}.

\begin{figure}[ht!]
 \includegraphics[width=0.9\columnwidth]{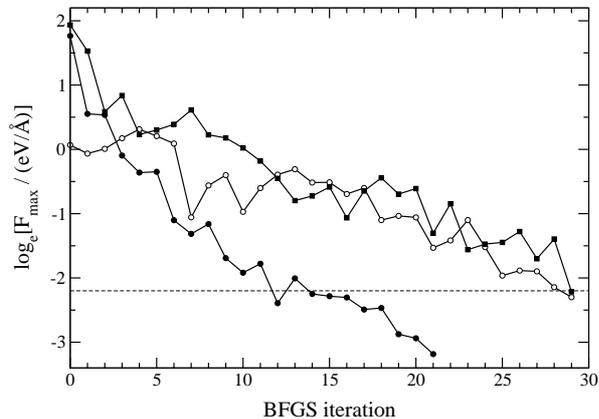}
\caption{\label{fig:bfgs-conv} Convergence behaviour of the maximum force
$|F_{\mathrm{max}}|$ on any atom as the BFGS algorithm proceeds for
the typical systems shown in Fig.~\ref{fig:bfgs-conv-pics}. CHAPS
Zwitterion (open circles); Al vacancy in alumina (filled circles);
GaAs nanocrystal (squares). Dashed line shows convergence threshold
of 0.11~eV/\AA{}. Simultaneous convergence of total energy and the
maximum displacement were required, with $|\mathrm{d}\mathbf{R}_{\mathrm{max}}|=0.002$~\AA{}\,
for 2 successive iterations was also required. From the starting coordinates
(see text) convergence was achieved in 29, 21 and 29 iterations respectively.}

\end{figure}

Fig.~\ref{fig:bfgs-conv} shows the convergence behaviour of the
maximum force as the BFGS algorithm progresses in each case. In all
three cases, convergence is achieved after 20-30 iterations. The forces
agree to good precision with those obtained in comparable calculations
in \textsc{castep}, so the optimisations follow a similar path. The
demands of convergence tolerance on plane-wave cut-off and sizes of
the localisation regions are not significantly greater than those
required for accurate evaluation of the energy in these systems. As
with plane-wave calculations, tight convergence of the electronic
energy is required before the forces are well-converged, since the
error in the forces scales approximately as the square root of the
error in the energy. We therefore conclude that it is possible to
perform geometry optimisation in the current framework with a similar
relative performance overhead compared to single-point energies as
in plane-wave DFT.

\subsection{Scaling with System Size}

\begin{figure}[ht]
\includegraphics[width=0.85\columnwidth]{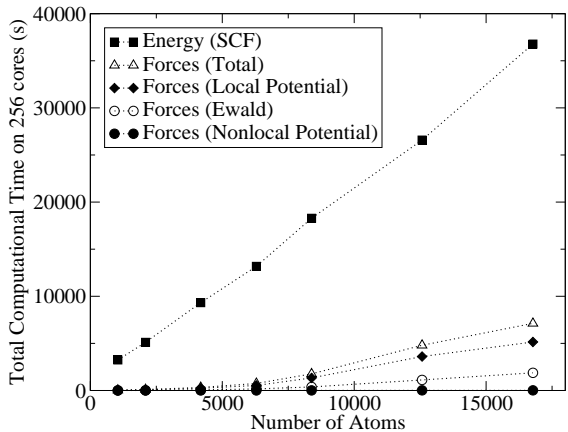} \caption{\label{fig:forces_DNA} Scaling of total computational time for SCF
total energy calculation for a series of DNA molecules of increasing
length (squares), compared to scaling of total computational time
for forces calculation (triangles), and the three force components
present: Ewald, Local Potential and Nonlocal Potential (diamonds,
empty circles and filled circles respectively), for the same simulations.}

\end{figure}

Finally, we demonstrate the scaling of the timings of the evaluation
of the forces compared to the total energy minimisation. As we have
described, the efficient parallel algorithms used ensure that despite
the $O(N^{2})$ prefactor on parts of the force calculation, the total
computational time remains dominated by optimisation of the NGWFs
and density kernel at each BFGS trial step up to very large $N$.
We show in Fig.~\ref{fig:forces_DNA} the total time taken by various
parts of the calculation for a series of systems each comprising double
helices of DNA of increasing length (with randomly chosen base pair
sequences to ensure no advantage can be gained through periodicity).
The base-pair sequences were generated randomly, and the atom positions
created with the Nucleic Acid Builder \cite{NAB} code. The positions
were relaxed within an empirical potential framework, using the Amber
code \cite{amber_2010}. This generated a starting point where the
forces on the atoms were low but non-zero, since the empirical-potential
forces do not exactly match those from the (presumably more accurate)
DFT calculation.

This system has previously been shown to exhibit good linear-scaling
with number of atoms $N$, and scale well to large numbers of processors\cite{hine-jcp-2010}.
The calculations here were performed with a plane-wave energy cut-off
of 700~eV, localisation radii 3.7~\AA{}, and a density kernel cutoff
radius of 16~\AA{}. These are somewhat lower accuracy values than
used in the previous tests, so as to allow scaling to very large system
sizes within moderate memory requirements, but should still allow
for reasonable convergence of the forces according to the findings
in Sec~\ref{sub:convergence_molecules}.\textbf{ }Note that in DNA,
with a very small HOMO-LUMO gap, the density matrix is quite long-ranged
and a relatively large cutoff must be used. 

We report in Fig.~\ref{fig:forces_DNA} timings for a total energy
minimisation followed by a calculation of the forces on 256 parallel
cores (Intel COREi7 CPUs). We vary the size of the system from 1042
atoms (16 base pairs) up to 16775 atoms (256 base pairs), scaling
the unit cell commensurately along one direction. Note that even the
smallest of these systems would be beyond the feasible scope of conventional
PWP methods, given the size of the simulation cell. The total time
for the optimisation of the electronic degrees of freedom is seen
to scale nearly perfectly as $O(N)$, while the calculation of the
local pseudopotential forces scales as roughly $O(N^{2})$ (though
the improved computational load balance at large system sizes masks
this slightly). Consequently, the fraction of the total time accounted
for by the forces increases from under 1\% to nearly 20\%. Eventually,
the calculation of forces would dominate and the method could no longer
be termed linear-scaling. However, this is not expected to be the
case in typical systems such as the DNA strands shown here until upwards
of 30000 atoms. Of course, this does not imply that a fully converged
geometry optimisation would be necessarily possible in linear-scaling
computational effort. An additional problem is the fact that ionic
relaxation requires a number of iterations, or evaluations of the
potential energy surface (PES), that increases with the number of
atoms\cite{Goedecker2001}. Preliminary steps have been taken by other
authors~\cite{Goedecker2001,Fernandez2003,nemeth-jcp04,nemeth-jcp05}
towards addressing this issue in the context of large-scale calculations.
This is outside the scope of the present work, however; often in practice
it will be sufficient to relax a smaller sub-region of a very large
system, thereby making the optimisation procedure tractable, or one
might in any case be investigating the effect of a localised perturbation
to an otherwise relaxed system. In such cases, we have shown that
geometry optimisation with plane-wave accuracy and linear-scaling
computational effort is achievable up to tens of thousands of atoms.

\section{Conclusions\label{sec:conclusions}}

In conclusion, we have shown that the combination of using strictly-localised
non-orthogonal generalised Wannier functions that move with the ions
but that are optimised on-the-fly within a basis set of psinc functions
that are fixed with respect to the ions, within the \textsc{onetep}
linear scaling DFT method, results in potential energy surfaces that
are sufficiently smooth that ionic forces can be calculated with high
accuracy. We have demonstrated that these forces can be systematically
converged with respect to energy cutoff and local orbital radius to
high-precision with low overhead relative to the demands of a comparable
total energy calculation.\textbf{ }We have demonstrated this by performing
geometry optimisation on a set of widely varied systems and comparing
to calculations using conventional plane-wave DFT. We note that for
weaker bonds, such as hydrogen bonds, although the discrepancy between
the PES and the calculated forces becomes more noticeable (Fig.~\ref{fig:water-dimer}),
the effect is nonetheless very small. Finally, we have demonstrated
that geometry optimisation is possible with a comparable computational
overhead to that for PWP simulations, and that the forces calculation,
while scaling as $\mathcal{O}(N^{2})$, remains a small fraction of
the total computational time until upwards of 30000 atoms for typical
systems.
\begin{acknowledgments}
We would like to thank Matt Probert for use of his BFGS algorithm.
The authors acknowledge the support of the Engineering and Physical
Sciences Research Council EPSRC Grant No. EP/G055882/1 for funding
through the HPC Software Development program. P.D.H. and C.-K.S. acknowledge
the support of University Research Fellowships from the Royal Society.
A.A.M. acknowledges the support of the RCUK fellowship program. The
authors are grateful for the computing resources provided by Imperial
College's High Performance Computing services CX1 and CX2, which have
enabled most of the simulations presented here. This work also made
use of the facilities of HECToR, the UK's national high-performance
computing service. HECToR is provided by UoE HPCx Ltd at the University 
of Edinburgh, Cray Inc and NAG Ltd, and funded by the Office of Science
and Technology through EPSRC's High End Computing Programme.
\end{acknowledgments}
\bibliographystyle{apsrev4-1}


%\newpage
%\printtables
%\newpage
%\noindent \textbf{Figure captions}
%\newpage
%\noindent \textbf{Caption for Figure .} 
%\newpage
%\printfigures

\end{document}